**Title page**

Title: A review of available software for adaptive clinical trial design

Running head: A review of adaptive design software

Authors: Michael J Grayling[1,2] and Graham M Wheeler[3]

Institutional affiliations:

1. MRC Biostatistics Unit, University of Cambridge, Cambridge, UK
2. Institute of Health & Society, Newcastle University, Newcastle upon Tyne, UK
3. Cancer Research UK & UCL Cancer Trials Centre, University College London, London, UK

Corresponding author: Michael J Grayling. Address: Institute of Health & Society, Newcastle University, Baddiley-Clark Building, Richardson Road, Newcastle upon Tyne NE2 4AX, UK. Email: michael.grayling@newcastle.ac.uk. Telephone: +44(0)191 208 7045

Grant support: This work was supported by the Medical Research Council [grant number MC_UU_00002/3 to MJG] and Cancer Research UK.

**Abstract**

**Background/Aims:** The increasing expense of the drug development process has seen interest in the use of adaptive designs (ADs) grow substantially in recent years. Accordingly, much research has been conducted to identify potential barriers to increasing the use of ADs in practice, and several articles have argued that the availability of user-friendly software will be an important step in making ADs easier to implement. Therefore, in this paper we present a review of the current state of software availability for AD.

**Methods:** We first review articles from 31 journals published in 2013-17 that relate to methodology for adaptive trials, in order to assess how often code and software for implementing novel ADs is made available at the time of publication. We contrast our findings against these journals' current policies on code distribution. Secondly, we conduct additional searches of popular code repositories, such as CRAN and GitHub, to identify further existing user-contributed software for ADs. From this, we are able to direct interested parties towards solutions for their problem of interest by classifying available code by type of adaptation.

**Results:** Only 29% of included articles made their code available in some form. In many instances, articles published in journals that had mandatory requirements on code provision still did not make code available. There are several areas in which available software is currently limited or saturated. In particular, many packages are available to address group sequential design, but comparatively little code is present in the public domain to determine biomarker-guided ADs.

**Conclusions:** There is much room for improvement in the provision of software alongside AD publications. Additionally, whilst progress has been made, well-established software for various types of trial adaptation remains sparsely available.

## Introduction

Classically, clinical trials have used fixed-sample designs. In this approach, a trial is designed, carried out using the design, and then the acquired data are analyzed on trial conclusion. In recent years, however, stagnation in the number of products submitted for regulatory approval (Hay *et al*, 2014), along with the escalating costs of drug development (DiMasi *et al*, 2016), has led the clinical trials community to seek new solutions to improving the efficiency of clinical research. One suggestion that has received much attention, is that a break from fixed-sample designs is required in order to make use of the pertinent information that is acquired over the course of a trial's progress. That is, to more regularly employ adaptive designs (ADs), which permit data-dependent modifications to be made to a trial's conduct through a series of prospectively planned interim analyses of the accumulating data. Indeed, both the US Food and Drug Administration and the European Medicines Agency have recognized that ADs could be key to the future of drug development (European Medicines Agency, 2007; US Food and Drug Administration, 2018).

With this increased interest, there has been an expansion in the publication of statistical methodology that facilitates the AD of clinical trials. Methods for carrying out numerous types of adaptation are now available, e.g., to drop particular treatment arms from further recruitment, to refine a trial's sample size, or to restrict future allocation to particular patient sub-groups. For an overview, several monographs have been published (Chow and Chang, 2011; Wassmer and Brannath, 2016; Yin, 2013). Furthermore, guidance is now available on when and why ADs may be useful, as well as on how to run such studies (Korn and Freidlin, 2017; Pallmann *et al*, 2018; Thorlund *et al*, 2018). Recommendations on how to report adaptively designed clinical trials are also under development (Dimairo *et al*, 2018).

However, the actual number of trials that have used ADs remains small. A recent review by Hatfield *et al* (2016) of phase II, phase II/III, and phase III trials registered on ClinicalTrals.gov between 29 February 2000 and 1 June 2014, along with trials from the National Institute for Health Research register, identified only 143 AD clinical trials. Similarly, Bothwell *et al* (2018) reviewed articles from several databases published prior to September 2014, and found 142 AD phase II, phase II/III, or phase III trials.

Accordingly, much research has also been conducted in order to identify and describe the potential barriers to the increased use of ADs (Chow and Corey, 2011; Cofey *et al*, 2012; Dimairo *et al*, 2015a; Dimairo *et al*, 2015b; Jaki, 2013; Kairalla *et al*, 2012; Meurer *et al*, 2016; Morgan *et al*, 2014; Quinlan *et al*, 2010; Love *et al*, 2017). Numerous barriers have since been identified, such as a lack of available expertise in AD, the requisite length of time required for trial design when using an AD, a fear AD would introduce operational biases, and inadequate funding structures. Here, our focus is on an additional barrier, which has been noted by several of these reviews: a lack of easily accessible, well-documented, user-friendly software for AD (Chow and Corey, 2011; Cofey *et al*, 2012; Dimairo *et al*, 2015a; Dimairo *et al*, 2015b; Jaki, 2013; Kairalla *et al*, 2012; Quinlan *et al*, 2010). The provision of software for ADs is particularly important because, relative to fixed-sample designs which often require only simple calculations, the complexity of ADs makes computational investigation of such methods typically a necessity. With the proliferation of software, it has been argued, project teams around the globe will be empowered to compare and contrast different designs in order to make informed decisions about the most appropriate design for their trial, and ultimately the frequency of appropriate AD use will increase. There have been recommendations that, wherever possible, software for novel AD methodology should be made available alongside statistical publications (Dimairo *et al*, 2015b).

Fortunately, therefore, several reviews of available software for ADs have now been presented. Zhu *et al* (2011) provided an overview of software for group sequential design, whilst Timofeyev (2014), Wassmer and Brannath (2016), and Wassmer and Vandemeulebroecke (2006) all provided more general overviews of software for ADs. However, each of these has concentrated on describing *what* software is available, focusing on established packages from a high-level perspective, and giving particular attention to stand-alone proprietary solutions such as East and AddPlan.

Here, our focus is directed toward two different aims. The first is to investigate the provision of user-contributed code and software for designing, conducting and analyzing trials using ADs in scientific publications. We review articles from a variety of journals that publish AD methodology, assessing how often code/software are provided alongside publications, and how these results compare with the current policies of these journals. Secondly, we assess which AD features are supported by available user-contributed programs for use in R, SAS, Stata, and other programming languages that are popular in the trials community. Since the abundance of user-written code makes it challenging to keep track of available solutions to particular design and analysis problems, we review several databases, including CRAN, SSC, and GitHub, in order to identify which design features and trial phases have been addressed heavily, and those that may require further computing resources to be provided.

We proceed by describing the methods behind our literature review, before detailing our findings on provision of code alongside AD methodology publications. We then detail identified available solutions in R, SAS, and Stata by type of adaptation, before discussing the current state of software for the AD of clinical trials.

## Methods

*Review protocol*

Here, we summarise the most important points behind our literature and repository review. Further details are given in the Supplementary Material.

*Review aims*

- To determine the frequency with which requisite computer code is made available alongside publications relating to the AD of clinical trials, and further classify this availability according to the archiving method and code completeness.
- To determine the most popular programming languages used within the AD community.
- To determine the degree to which authors who state computer code is "available upon request", are able to respond with said code following an e-mail request.
- To identify and describe user-written code relating to the AD of clinical trials, with a focus on R, SAS, and Stata.

*Identification of relevant journal publications*

*PubMed Central search*. PubMed Central was searched on July 5 2018 by MJG, in order to identify potential publications for inclusion in our review. Articles were required to have been published in one of 31 journals, a bespoke selection of those we believed to be most likely to publish articles relating to AD methodology (see Supplementary Table 1). Publications from each journal were identified by searching the [Abstract], [Body – Key Terms], and [Title] fields for 53 chosen AD-related terms, which are listed in the Supplementary Material. The search was limited to those articles published between January 1 2013 and December 31 2017. Supplementary Table 1 provides the number of records identified for each of the considered journals; in total 4123 articles were identified for review.

*Publication inclusion criteria*. We desired to include publications related to the design and analysis of AD clinical trials. Thus, using the US Food and Drug Administration definition of an AD (US FDA, 2018), our inclusion criteria were:

1. A publication that proposes or examines design or analysis methodology for a clinical that "allows for prospectively planned modifications to one or more aspects of the design based on accumulating data from subjects in the trial" (US FDA, 2018);
2. A complete peer-reviewed publication (i.e., we excluded conference abstracts);
3. Set within the context of clinical trials (i.e., we excluded methodology that could be used for the AD of a clinical trial if the primary motivation was not clinical trial research);
4. Performs computational work of any kind relating to ADs (i.e., even to confirm theoretical results, produce simple graphs, etc.).

Note that we excluded conference abstracts as we believed it would be unlikely that they would explicitly note whether/where code is available. Similarly, in fields other than clinical trials there may be different expectations on the availability of code. We thus excluded such publications to reduce the bias in our findings, given our primary interest was AD methodology for clinical trials. No restrictions were made on the level of code required for inclusion since we felt drawing such conclusions would be subjective. Finally, note that by criterion 1, we exclude publications that simply present the results of a clinical trial that utilized an AD.

*Selection of studies for inclusion in the review and data extraction*. Two-hundred records were randomly selected to pilot the selection process and data extraction upon. Specifically, MJG and GMW independently considered the 200 records for inclusion, and for each of those marked for inclusion, extracted the following data:

- Software availability: Each of the articles were allocated in to one of the categories given in Supplementary Table 2, according to the provision of the code required for the presented results.
- Software languages used: C++, R, SAS, Stata, Unclear, etc.

Following this pilot, areas of disagreement were discussed in order to enhance the reliability of the selection process and data extraction on the remaining 3923 records, which were allocated evenly and at random to MJG and GMW. In extreme cases where a reviewer was unable to come to a conclusion on inclusion/data extraction, a decision was made following discussion with the other reviewer.

Note that in each case of exclusion, a reason for exclusion amongst the following options was recorded:

- Non-adaptive design methodology;
- No code required;
- Not within the context of clinical trials;
- Not complete publication.

*Identification of relevant database-archived computer code*

*Software-specific database searches*. In order to identify further software for the AD of clinical trials that is available for R, SAS, and Stata, MJG performed the following additional software-specific database searches on 10 July 2018. For each, there was no simple means of extracting results data in to a manageable offline form. Therefore, a less formal approach to record identification had to be taken, as outlined below.

Firstly, Rseek was used to identify packages currently available on the Comprehensive R Archive Network (CRAN; the principal location for the storage of R packages) that are pertinent to ADs. Specifically, each of the 53 terms used in the article search of PubMed Central (the “search terms”) were entered in to the engine at https://rseek.org/. Next, the articles from the R-project tab were screened, with any that appeared to be of potential relevance to ADs noted in a .csv file. Similarly, to identify code available for Stata that is relevant to ADs, the Statistical Software Components (SSC) archive was used (which hosts the largest collection of user-contributed Stata programs). The search terms were entered in to the search bar at https://ideas.repec.org/. Any potentially germane results were added to the aforementioned .csv file. Moreover, the search terms were also entered in to the search engine at https://www.stata-journal.com/, in order to identify relevant publications in the *Stata Journal* (note that we did not search for *Stata Journal* articles via PubMed Central, as not all such articles are indexed there), the premier journal for the publication of Stata code articles. To find user-contributed code for AD in SAS, the abstracts of the proceedings of the SAS Global Forums from 2007-18 were searched using the search terms given earlier (e.g., for 2016 the terms were utilized via Ctrl+F searches at support.sas.com/resources/papers/proceedings16/). Finally, the procedure was repeated on GitHub, using the search bar at https://github.com/, with all seemingly relevant results again stored in a .csv file. For this search, no restrictions on the programming language utilized were made.

Note that for each of these databases, no limits on the publication date were employed, as our goal was to identify as much relevant software as possible. The number of records identified of potential relevance are given in Supplementary Table 3.

*Identification of relevant records*. Each of the records from the search described in Section 2.4.1 were screened in order to identify those related to ADs. Our criteria for listing a record as relevant was point 1 from Section 2.3.2. The functionalities of those that were relevant were also noted via a checklist, using one or more of the following keywords:

Adaptive randomization; Alpha spending; Bayesian methods; Biomarker-based methods; Dose-modification/escalation; Drop the loser; Group sequential; Multi-stage; Phase I; Phase I/II; Phase II; Phase II/III; Phase III; Pick the winner; Sample-size adjustment; Stopping rules; Two stage.

To pilot the screening, 31 records (~10% of the 307 records initially identified) were chosen at random and reviewed by MJG and GMW. As above, this allowed for discussions on differences of opinion, in order to improve the standardization of the classification for the remainder of the records. For efficiency purposes, MJG then screened each of the remaining records from GitHub. GMW screened those from each of the other databases.

Finally, note that for all records that were marked to be of relevance to ADs, the author's additional repositories were screened (e.g., via their homepage on GitHub) in order to identify any further code relating to ADs. From this, three previously unidentified records were included.

## Results

*Code availability*

Our search yielded 4,123 articles across the 31 considered journals (Supplementary Table 1). Of these, 3,875 were excluded on the following grounds: Non-adaptive design methodology (3,817); Not a complete peer-reviewed paper (40); Not directly applied to clinical trials (13); No computational work required (6). This left 247 eligible articles across 26 journals for our review.

Of these 247 articles, 144 (58.3%) did not provide code used for computational work. Thirty-two articles (13%) provided complete code in the article or its supplementary material to either recreate the exact outputs of the paper, or provided all functions to do so; a further 8 articles (3.2%) provided partial code. Twenty-seven articles (10.9%) provided URL addresses to websites where code was to be stored; of these 27, only 13 (48%) were accessible at the time of review, with the remainder either not providing the code for the relevant article, or the URLs no longer worked. Six articles (2.4%) stated that code was available in online supplementary material, but the code was not present. In another six articles (2.4%), code was either released as standalone software or a downloadable package for a specific program (4/6), or the functions were incorporated into an already-existing package (2/6). One article cited software that could be used for the purpose it outlined, but no further details were provided, and another used commercial software for their work entirely, so no code/instructions were provided.

The remaining 22 articles (8.9%) stated that code was available upon request from the corresponding author. For all 22 articles we sent request e-mails to the corresponding author, explaining that the article in question stated code was available upon request and that we were asking for it as part of a literature review on the availability of AD code (e-mail template given in the Supplementary Material). Authors were given one month from the date of e-mail to reply and were sent a reminder e-mail two weeks after first contact if they had not responded by this time. From these 22 requests, 14 (63.6%) authors replied to either provide the code used, or to direct us to a URL where the code was deposited; one author replied to say that the code was not available. Six authors (27.3%) did not reply to our request and one additional author was uncontactable via the stated corresponding author's e-mail address. A search for an up-to-date address for the corresponding author yielded no leads.

Incorporating the author responses that provided code or accessible URLs to code in to our results means that code was made available and accessible (either directly in the paper, via a valid URL, or incorporated into an available software package) for 65 articles (26.3%), with a further 8 articles (3.2%) providing partial code; the remaining 174 articles (70.5%) did not provide code relating to the proposed AD methodology. Figure 1 shows the distribution of code provision by journal.

*Code availability and journal policies*

Policies on whether computer code should be provided with article submissions vary between journals. In our analyses we reviewed journal policies on providing computer code and compared them to the observed rates of code availability/provision in our review. Figure 2 shows the distribution of code provision across journals according to their code provision policy (Compulsory, Strongly Encouraged, Encouraged, Possible, Not Mentioned). The data show journals with compulsory policies for code provision have not been enforcing their policies.

There is a possibility that articles published at the start of our review period (i.e., 2013) may not have been subject to the same code provision policy that is in place now. However, violations of the compulsory policy type are consistent across the review period. For example, *Statistics in Medicine* (ISSN 0277-6715; Wiley Online Library) states that "The journal also *requires* authors to supply any supporting computer code or simulations that allow readers to institute any new methodology proposed in the published article."; this is an example of a compulsory policy. In our review, 66 articles published in *Statistics in Medicine* were considered eligible. Table 1 shows the distribution of articles published each year across journal provision type for articles published in *Statistics in Medicine*. Over 5 years, 51 articles (77%) were published with no code provided, and numbers did not noticeably decrease over time, which would be consistent with the introduction of a compulsory code provision policy. On the other hand, *Biometrical Journal* (ISSN 1521-4036; Wiley Online Library) has a "Strongly Encouraged" level policy, stating "The journal strongly supports Reproducible Research. Authors are therefore vigorously encouraged to submit computer code and data sets used to illustrate new methods and to reproduce the results of the paper." All 7 eligible articles published during our review period provided full code to reproduce the results or gave all functions required to replicate the paper's results.

*Software used*

A variety of different statistical programs were used in the eligible articles, including open-source libraries, licensed programs, and commercial software. Overall, 129 articles (52%) stated what software was used in their computations; 60 of these articles (47%) did not make their code available.

Of the 129 articles, 107 used R (R Core Team, 2017); 91 such articles used R only, and the other 16 used R in combination with another program (e.g., MCMC sampling software such as JAGS, OpenBUGS, or WinBUGS) or provided code/software in other computing languages as well as R. Table 2 shows the usage of different software and their provision categories in journals.

*Repository review*

We performed additional searches of major software libraries to identify and classify available computer software related to ADs. Our searches found 310 software libraries, of which 123 were considered eligible. Of these records, 64 (52%) were found on CRAN; 45 of these 64 CRAN packages had duplicate repositories on GitHub pages. Forty (33%) additional repositories were found on GitHub (i.e., repositories not located on any other platform), 8 (7%) on SSC, 6 (5%) from the SAS Global Forum and 5 (4%) from the *Stata Journal*. Of the 40 GitHub repositories, 35 (88%) featured code for R; the remaining 5 entries featured code for Julia (2/40), Javascript (1/40), Python (1/40), and SAS (1/40). This means that of the 123 eligible repositories, 99 (80%) provided R packages, or code for use in R that is yet to be published as a package on CRAN.

Table 3 shows the primary applications for AD software, split by software language and intended trial phase. The majority of available packages cover phase II and phase III trials and are for group sequential methods. The packages/programs tended to cover multiple purposes; 64 programs belonged to one of the design categories listed in Table 3, 51 belonged to two categories, 7 belonged to three categories, and 1 covered four categories.

Supplementary Table 4 shows the distribution of software and trial phase catered for by the different subcategories of group sequential methods. When breaking down the "Group sequential" designs category into its constituent parts, we see that many packages are available for dealing with both two-stage and multi-stage designs. As per previous tables, R is generally the favoured software for writing such programs.

We also extracted, where possible, the date when the package was last updated or released. For 11 (9%) entries, only the year of last known update was available. Figure 3 shows the distribution of year of latest update by Repository. Most packages are hosted on CRAN and GitHub, repositories that users can easily update and submit packages to, and all CRAN packages have been released or updated within the last 4 years. There are few programs hosted on the SAS Global Forum, SSC, and via the *Stata Journal*, most of which have not been updated in the last 4 years. We cannot tell if the lack of updates are because the package is in perfect working order with all required functionality, or whether a lack of interest from users means there is no need for the maintainer to update it.

**Discussion**

By scanning 31 journals and five years' worth of publications, we provide reliable estimates of the prevalence of software provision alongside AD methodology publications. The reliability of our findings is also aided by joint-review of 10% of records, with discussion of findings to ensure consistency. Ultimately, we found that 71% of included articles did not provide any code or software. Most of the journals in which these articles were published have code provision policies that either require or strongly encourage the provision of code.

The low rate of software provision is a disappointing finding. Providing code alongside methodological research allows readers to reproduce novel ADs and tailor them to their own project needs. Some research funders expect funding recipients to make data and original software used for analyses fully available at the time of publication. For example, the Wellcome Trust state that researchers should make sure such outputs i) are discoverable, ii) use recognized community repositories, and iii) use persistent identifiers (e.g., DOIs) where possible (see https://wellcome.ac.uk/funding/guidance/policy-data-software-materials-management-and-sharing). We recommend that this guidance is followed for all AD related publications whenever feasible.

More positively, we identified that there has been a marked increase in the number of software repositories relating to ADs over the last five years (Figure 3). A further interesting result is that the majority of AD-related programs are written for R. Therefore, whilst provision of code and software with new publications may help increase the use of ADs, it would also thus be prudent for statisticians to be familiar with how to use R. Furthermore, by demonstrating what trial adaptations are covered by existing software, we have made it possible for researchers to be better informed as to where new and improved code is required. In particular, many programs are available for group sequential design. Future research in these areas likely does not require the provision of brand-new code, when several open-source packages are likely already available for the required AD. In contrast, only limited software is available to support sample size re-estimation, or biomarker-based adaptation.

A limitation of our review is that some papers may not release code at the time of publication as they intend to release their code as part of a larger package, or because of potential confidentiality issues. However, no papers mentioned that this was the case, and we would encourage authors to state why code is not available to accompany research.

In summary, to overcome the barriers to implementing ADs in clinical trials, we encourage researchers to make their code available alongside their published research as supplementary material, or by storing it on stable repositories such as GitHub and CRAN. Several articles stated code was available at a given URL, but half of these URLs did not work. Similarly, about a third of articles that stated code would be available upon request were unable to provide code within a month of sending a written request. Accordingly, making code available in either of these manners should not be viewed as a reliable long-term method of user access.

**Declaration of Conflicting Interests**

The Authors declare that there is no conflict of interest.

**Table 1.** Code provision for articles published in Statistics in Medicine, split by year of first publication.

| | Year of publication | | | | | |
|---|---|---|---|---|---|---|
| | 2013 | 2014 | 2015 | 2016 | 2017 | Total |
| Code not available | 4 | 12 | 13 | 12 | 10 | 51 |
| Full code/package provided/accessible | 1 | 3 | 2 | 3 | 5 | 14 |
| Partial code provided | 0 | 0 | 0 | 0 | 1 | 1 |
| Total | 5 | 15 | 15 | 15 | 16 | 66 |

**Table 2.** Software used in adaptive design articles (where stated) across code provision category.

| | Code not available | Full code/package provided/accessible | Partial code provided | Total |
|---|---|---|---|---|
| C | 1 | 2 | - | 3 |
| Excel | - | 2 | - | 2 |
| FACTS | 2 | 1 | - | 3 |
| FORTRAN | 1 | 2 | - | 3 |
| JAGS | 1 | - | 1 | 2 |
| Matlab | - | 2 | - | 2 |
| PASS | 1 | - | - | 1 |
| R[a] | 52 | 49 | 6 | 107 |
| EAST | - | - | 1 | 1 |
| SCPRT | - | - | 1 | 1 |
| OpenBUGS | 1 | - | - | 1 |
| WinBUGS | 1 | 1 | 3 | 5 |
| Stand-alone program | 1 | 7 | - | 8 |
| P3M | 1 | - | - | 1 |
| Stata | 2 | 2 | 1 | 5 |
| SAS | 2 | 3 | - | 5 |

[a]Includes custom R functions, use of existing R packages, and also R Shiny applications.

**Table 3.** Main Functions of software repositories, split by software and trial phase. Each package may belong to multiple categories and cover multiple trial phases.

| | | Group sequential methods[a] | Dose modification / escalation | Sample size adjustment | Adaptive randomization | Bayesian methods | Biomarker-based methods |
|---|---|---|---|---|---|---|---|
| Software | JavaScript | - | 1 | - | - | - | - |
| | Julia | 1 | - | 1 | 1 | - | 1 |
| | Python | 1 | - | - | - | 1 | - |
| | R | 62 | 36 | 6 | 7 | 45 | 9 |
| | SAS | 6 | 1 | - | 4 | 1 | 1 |
| | Stata | 11 | 1 | 1 | - | 4 | - |
| | | | | | | | |
| Phase | I | 2 | 27 | - | - | 21 | - |
| | I/II | 2 | 10 | - | - | 8 | 1 |
| | II | 75 | 10 | 8 | 12 | 27 | 10 |
| | II/III | 3 | 1 | - | - | 1 | - |
| | III | 57 | 2 | 6 | 11 | 15 | 7 |

[a]"Group sequential methods" covers the following subcategories: group sequential; two-stage; multi-stage; stopping rules; drop the loser; pick the winner; alpha spending.

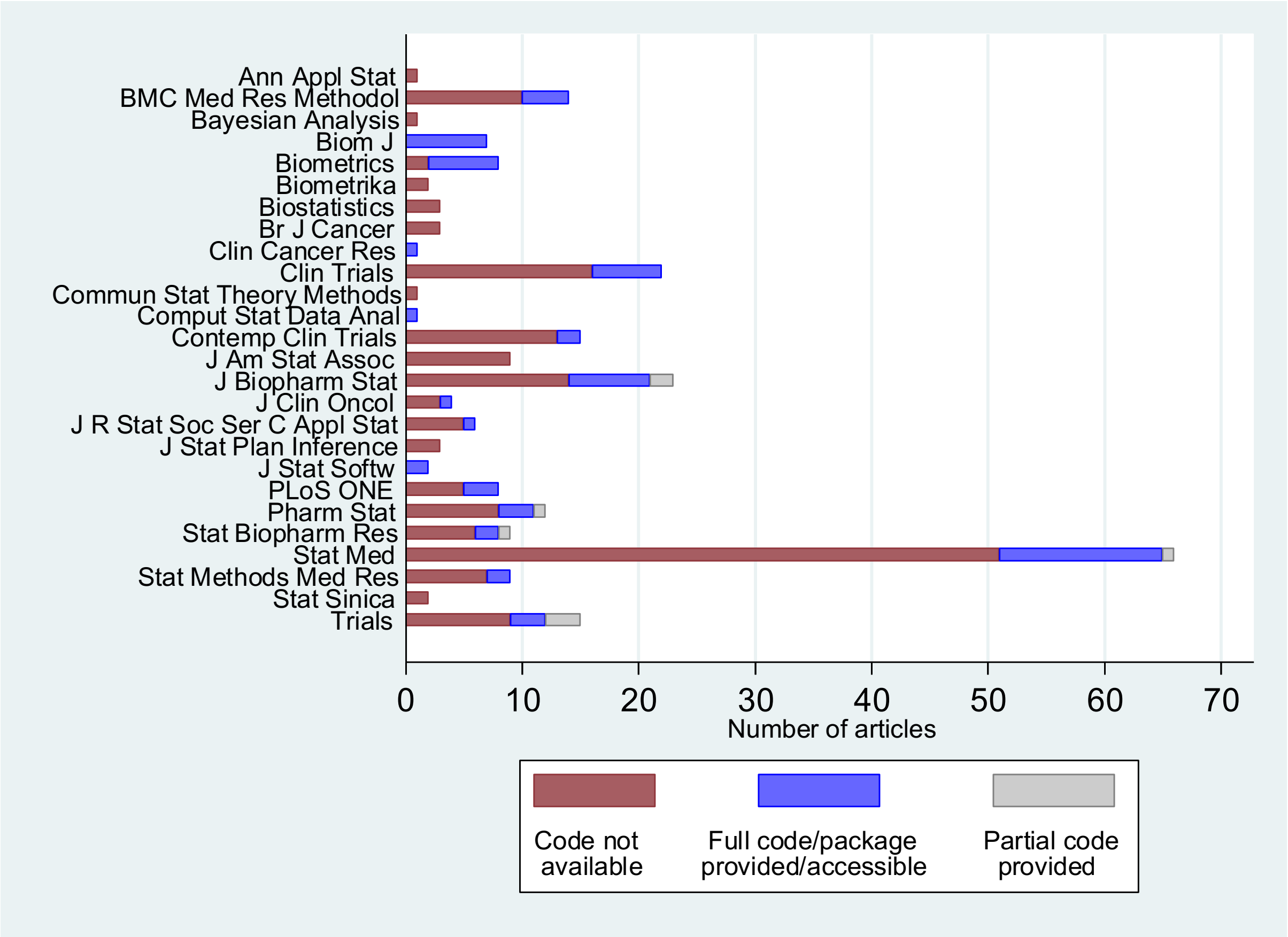

**Figure 1.** Number of articles by journal and whether code is provided or not.

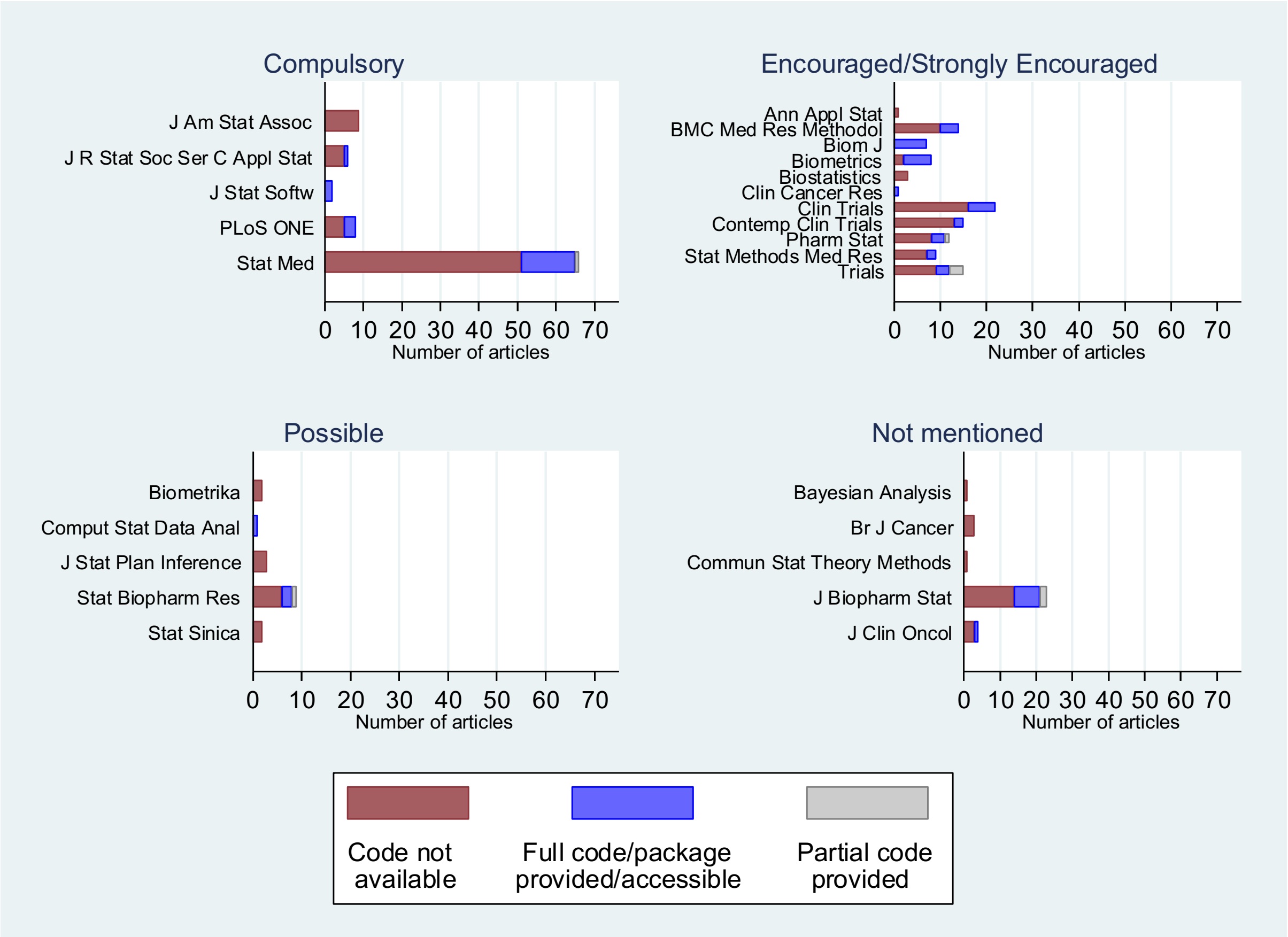

**Figure 2.** Number of articles by code provision, journal and journal's code provision policy.

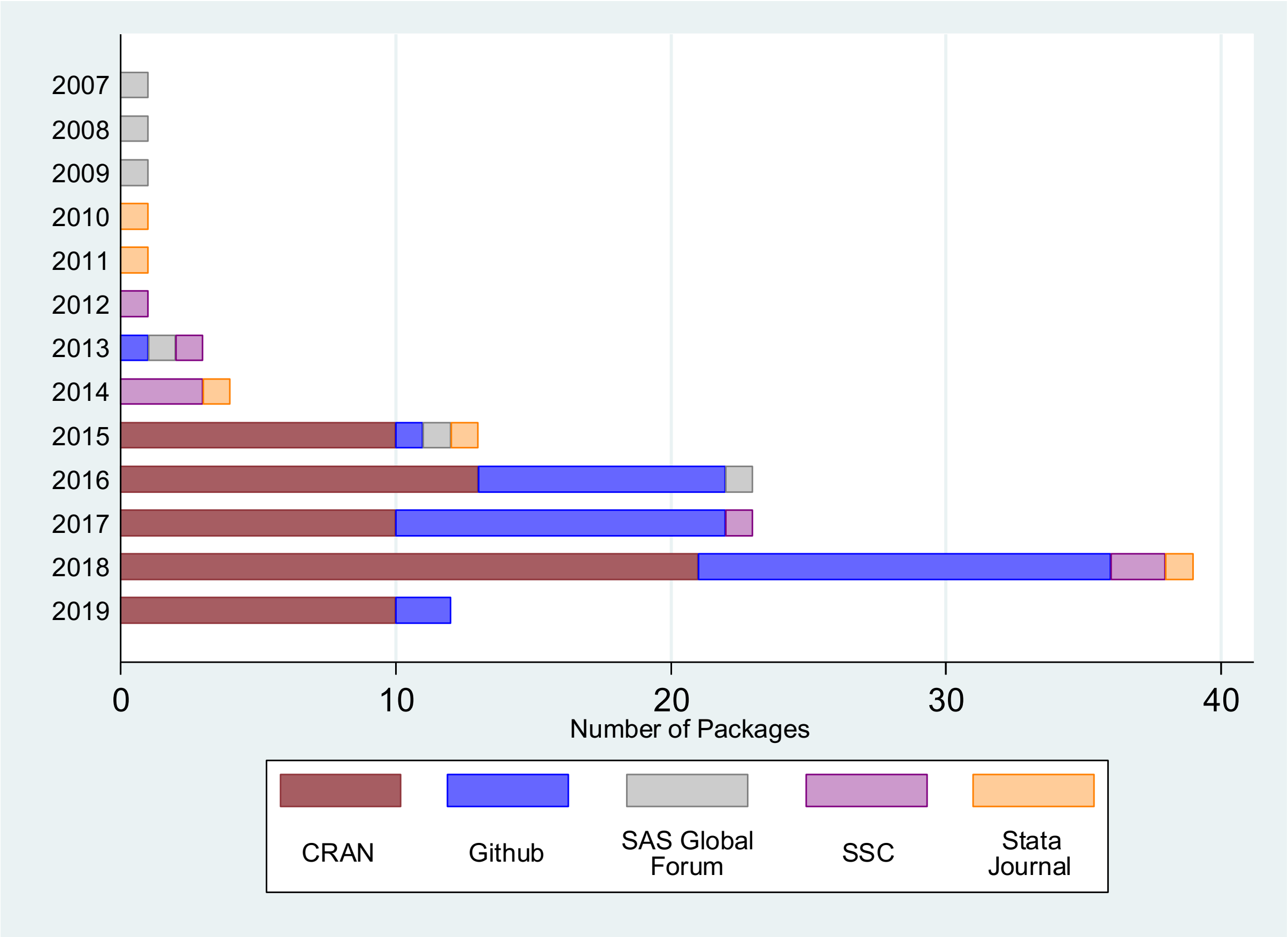

**Figure 3.** Number of identified repositories by location and year.

**Supplementary Material**

*Pubmed Central search*

The following 53 search terms were used in relation to the PubMed Central search described in Section 2.3.1 of the main manuscript

Adaptive approach; adaptive clinical; adaptive design; adaptive method; adaptive randomisation; adaptive randomization; adaptive rule; adaptive trial; alpha spending; bayesian adaptive; bayesian approach; bayesian clinical; bayesian design; bayesian method; bayesian model; bayesian rule; bayesian trial; dose-escalation; dose-selection; dose escalation; dose finding; dose selection; drop the loser; flexible design; group sequential; interim analysis; multi-stage; multi stage; phase 1; phase 1-2; phase 1/2; phase 2; phase 2-3; phase 2/3; phase 3; phase I; phase I-II; phase I/II; phase II; phase II/III; phase III; pick the winner; play the winner; response adaptive; sample adjust; sample re-estimation; sample reestimation; sample size adjust; sample size re-estimation; sample size reestimation; stopping rule; two-stage; two stage.

Note that the following term-field combinations were found to be absent from PubMed Central, and so were subsequently omitted from the search text adaptive clinical[Abstract]; bayesian clinical[Abstract]; bayesian trial[Abstract]; sample adjust[Abstract]; sample re-estimation[Abstract]; sample reestimation[Abstract]; sample size adjust[Abstract]; bayesian clinical[Body - Key Terms]; drop the loser[Body - Key Terms]; phase 2/3[Body - Key Terms]; phase II-III[Body - Key Terms]; pick the winner[Body - Key Terms]; sample adjust[Body - Key Terms]; sample re-estimation[Body - Key Terms]; adaptive clinical[Title]; adaptive

randomisation[Title]; bayesian trial[Title]; bayesian clinical[Title]; sample re-estimation[Title]; sample size adjust[Title]; sample adjust[Title]; play the winner[Title]; sample reestimation[Title].

Thus, as an example, the extracted records for the *Journal of Statistical Software* were found with the following search

((((((((((((((((((((((((((((((((((((((((((((((((((((((((((((((((((((((((((((((((((((((((((((((((((((((((((((((((((((((((((((((((((((((((((((((((((((((multi-stage[Abstract]) OR multi-stage[Body - Key Terms]) OR multi-stage[Title]) OR multi stage[Abstract]) OR multi stage[Body - Key Terms]) OR multi stage[Title]) OR adaptive design[Abstract]) OR adaptive trial[Abstract]) OR adaptive rule[Abstract]) OR adaptive method[Abstract]) OR adaptive approach[Abstract]) OR adaptive randomisation[Abstract]) OR adaptive randomization[Abstract]) OR flexible design[Abstract]) OR group sequential[Abstract]) OR bayesian design[Abstract]) OR bayesian adaptive[Abstract]) OR bayesian model[Abstract]) OR bayesian approach[Abstract]) OR bayesian rule[Abstract]) OR bayesian method[Abstract]) OR two stage[Abstract]) OR interim analysis[Abstract]) OR sample size re-estimation[Abstract]) OR stopping rule[Abstract]) OR drop the loser[Abstract]) OR pick the winner[Abstract]) OR play the winner[Abstract]) OR dose selection[Abstract]) OR dose-selection[Abstract]) OR dose escalation[Abstract]) OR dose-escalation[Abstract]) OR dose finding[Abstract]) OR phase 2-3[Abstract]) OR phase 2/3[Abstract]) OR phase II-III[Abstract]) OR phase II/III[Abstract]) OR phase 1-2[Abstract]) OR phase 1/2[Abstract]) OR phase I-II[Abstract]) OR phase I/II[Abstract]) OR alpha spending[Abstract]) OR response adaptive[Abstract]) OR phase I[Abstract]) OR phase II[Abstract]) OR phase III[Abstract]) OR phase 1[Abstract]) OR phase 2[Abstract]) OR phase 3[Abstract]) OR adaptive design[Body - Key Terms]) OR adaptive trial[Body - Key Terms]) OR adaptive clinical[Body - Key Terms]) OR adaptive rule[Body - Key Terms]) OR adaptive method[Body - Key Terms]) OR adaptive approach[Body - Key Terms]) OR adaptive randomisation[Body - Key Terms]) OR adaptive randomization[Body - Key Terms]) OR flexible design[Body - Key Terms]) OR group sequential[Body -

Key Terms]) OR bayesian design[Body - Key Terms]) OR bayesian adaptive[Body - Key Terms]) OR bayesian model[Body - Key Terms]) OR bayesian approach[Body - Key Terms]) OR bayesian trial[Body - Key Terms]) OR bayesian rule[Body - Key Terms]) OR bayesian method[Body - Key Terms]) OR two stage[Body - Key Terms]) OR interim analysis[Body - Key Terms]) OR sample size re-estimation[Body - Key Terms]) OR sample size adjust[Body - Key Terms]) OR stopping rule[Body - Key Terms]) OR play the winner[Body - Key Terms]) OR dose selection[Body - Key Terms]) OR dose-selection[Body - Key Terms]) OR dose escalation[Body - Key Terms]) OR dose-escalation[Body - Key Terms]) OR dose finding[Body - Key Terms]) OR phase 2-3[Body - Key Terms]) OR phase II/III[Body - Key Terms]) OR phase 1-2[Body - Key Terms]) OR phase 1/2[Body - Key Terms]) OR phase I-II[Body - Key Terms]) OR phase I/II[Body - Key Terms]) OR alpha spending[Body - Key Terms]) OR response adaptive[Body - Key Terms]) OR phase 1[Body - Key Terms]) OR phase 2[Body - Key Terms]) OR phase 3[Body - Key Terms]) OR phase I[Body - Key Terms]) OR phase II[Body - Key Terms]) OR phase III[Body - Key Terms]) OR adaptive design[Title]) OR adaptive trial[Title]) OR adaptive rule[Title]) OR adaptive method[Title]) OR adaptive approach[Title]) OR adaptive randomization[Title]) OR flexible design[Title]) OR group sequential[Title]) OR bayesian design[Title]) OR bayesian adaptive[Title]) OR bayesian model[Title]) OR bayesian approach[Title]) OR bayesian rule[Title]) OR bayesian method[Title]) OR two stage[Title]) OR interim analysis[Title]) OR sample size re-estimation[Title]) OR stopping rule[Title]) OR drop the loser[Title]) OR pick the winner[Title]) OR dose selection[Title]) OR dose-selection[Title]) OR dose escalation[Title]) OR dose-escalation[Title]) OR dose finding[Title]) OR phase 2-3[Title]) OR phase 2/3[Title]) OR phase II-III[Title]) OR phase II/III[Title]) OR phase 1-2[Title]) OR phase 1/2[Title]) OR phase I-II[Title]) OR phase I/II[Title]) OR alpha spending[Title]) OR response adaptive[Title]) OR phase 1[Title]) OR phase 2[Title]) OR phase 3[Title]) OR phase I[Title]) OR phase II[Title]) OR phase III[Title]) OR two-stage[Abstract]) OR two-stage[Body - Key Terms]) OR two-stage[Title]) OR sample size reestimation[Abstract]) OR sample size reestimation[Body - Key Terms]) OR sample size reestimation[Title]) OR sample reestimation[Body - Key Terms]) AND ("2013/01/01"[Publication Date] : "2017/12/31"[Publication Date])) AND "Journal of Statistical Software"[Journal]))

For those articles which were included, Supplementary Table 2 then provides further information on the possible classification of the level of code provision.

*Code Provision Policies by Journal*

Please contact the corresponding author for a spreadsheet containing further details on the Journal policies on code provision.

*E-mail template for code requests*

**Subject:** Request for computer code: <authors> (<journal>, <year>)

**Body:**

Dear <author>,

We are two UK researchers conducting a survey on the provision of software and computer code in academic publications, specifically focusing on adaptive designs in clinical trials. One aim of this work is to assess the availability of software and code for researchers in adaptive designs, and how complete any available code is.

One of the eligible studies in our survey was the paper "<article title>" (<authors>, <journal>, <volume>(<issue>):<pages>. <year>), for which you are the corresponding author. In this article, it says <text stating code is available upon request>. For eligible articles that say code is available upon request, we are e-mailing the corresponding author to request the code used.

Would you be able to provide us with the code used to generate the results and figures in this paper? If so, please send all relevant and available code files to Dr. Graham Wheeler (graham.wheeler@ucl.ac.uk) by <date one month from date e-mail is to be sent> at the latest.

If it is not possible to provide us with any of the code files used, please do reply informing us of this.

Yours sincerely,

Dr. Graham Wheeler, CRUK and UCL Cancer Trials Centre, University College London, UK (graham.wheeler@ucl.ac.uk)
Dr. Michael Grayling, Institute of Health and Society, University of Newcastle, UK (michael.grayling@newcastle.ac.uk)

*Database searches*

Supplementary Table 3 provides information on the number of records identified in each of the considered databases, whilst Supplementary Table 4 shows the distribution of software and trial phase catered for by the different subcategories of group sequential methods.

**Supplementary Table 1.** A list of the journals included in the search procedure, and the number of identified potential records for inclusion for each, is given.

| Journal | Records |
|---|---|
| *Annals of Statistics* | 4 |
| *Bayesian Analysis* | 10 |
| *Biometrical Journal* | 12 |
| *Biometrics* | 62 |
| *Biometrika* | 8 |
| *Biostatistics* | 21 |
| *BMC Medical Research Methodology* | 71 |
| *British Journal of Cancer* | 276 |
| *Clinical Cancer Research* | 282 |
| *Clinical Trials* | 55 |
| *Communications in Statistics: Theory and Methods* | 2 |
| *Computational Statistics & Data Analysis* | 6 |
| *Contemporary Clinical Trials* | 64 |
| *Journal of Biopharmaceutical Statistics* | 35 |
| *Journal of Clinical Oncology* | 288 |
| *Journal of Statistical Planning and Inference* | 6 |
| *Journal of Statistical Software* | 3 |
| *Journal of the American Statistical Association* | 41 |
| *Journal of the Royal Statistical Society. Series B, Statistical Methodology* | 4 |
| *Journal of the Royal Statistical Society. Series C, Applied Statistics* | 20 |
| *Pharmaceutical Statistics* | 14 |
| *PLoS ONE* | 2094 |
| *Statistica Sinica* | 7 |
| *Statistical Methods in Medical Research* | 41 |
| *Statistics & Probability Letters* | 1 |
| *Statistics in Biopharmaceutical Research* | 14 |
| *Statistics in Medicine* | 154 |
| *Technometrics* | 1 |
| *The Annals of Applied Statistics* | 25 |
| *The BMJ* | 26 |
| *Trials* | 476 |
| Total | 4123 |

**Supplementary Table 2.** Possible classifications on the provision of software of publications included in the review, and a more detailed description of their meaning, is provided.

| Classification | Meaning |
|---|---|
| Included in paper/appendix – Complete code provided (to exactly reproduce results) | The relevant code is made available with the publication, either as part of the main paper or in an (potentially online) appendix. This code is an executable form that allows the results from the paper to be reproduced exactly |
| Included in paper/appendix – Complete code provided (functions only) | The relevant code is made available with the publication, either as part of the main paper or in an (potentially online) appendix. All functions that are in theory required to be able to reproduce the results from the paper are supplied, but not in an executable form for the exact production of results |
| Included in paper/appendix – Partially complete code provided | Relevant code is made available with the publication, either as part of the main paper or in an (potentially online) appendix. However, the code does not allow for all of the calculations in the publication to be replicated |
| Incorporated into previously released software/package | The relevant code is included as a new feature in a previously released software package |
| Released as stand-alone software/package | The relevant code is released as a stand-alone, referenced, and typically downloadable, software package |
| URL to website provided (still accessible) | A URL is given to where relevant code could be acquired, and the code remains accessible at this address |
| URL to website provided (no longer accessible) | A URL is given to where relevant code could be acquired. However, the code can no longer be accessed at this address |
| Made available upon request | Code is listed as being available upon e-mail request from one of the publication's authors |
| Not discussed/made available | The provision of relevant code is not discussed or made available in any of the forms given above |
| Other | Several articles did not fall in to any of the above classifications. A detailed description of why is provided in the main article |

**Supplementary Table 3.** The number of identified potential records of relevance to the AD of clinical trials, from our search of software-specific database searches, is given.

| Database | Records |
|---|---:|
| GitHub | 161 |
| RSeek | 103 |
| SAS Global Forum | 16 |
| SSC | 9 |
| *Stata Journal* | 21 |
| Total | 310 |

**Supplementary Table 4.** Group sequential designs broken down into constituent parts by software.

| | Group sequential | Two-stage | Multi-stage | Stopping Rules | Drop the loser | Pick the winner | Alpha spending |
|---|---|---|---|---|---|---|---|
| JavaScript | - | - | - | - | - | - | - |
| Julia | 1 | 1 | - | 1 | - | - | - |
| Python | 1 | - | 1 | 1 | - | - | - |
| R | 45 | 33 | 30 | 34 | 2 | 3 | 11 |
| SAS | 3 | 2 | 4 | 2 | - | - | 1 |
| Stata | 8 | 3 | 7 | 9 | 1 | 1 | 1 |